# Fourier Synthetic Aperture-based Time-resolved Terahertz Imaging

Vivek Kumar[1,*], Pitambar Mukherjee[2], Lorenzo Valzania[1], Amaury Badon[3], Patrick Mounaix[2], Sylvain Gigan[1]

[1]Laboratoire Kastler Brossel, ENS-Université PSL, CNRS, Sorbonne Université, Collège de France, 24 rue Lhomond, 75005 Paris, France.
[2]IMS Laboratory, University of Bordeaux, UMR CNRS 5218, 351 Cours de la Libération Bâtiment A31, 33405 Talence, France.
[3]Laboratoire Photonique Numérique et Nanosciences (LP2N), UMR 5298, University of Bordeaux, F-33400 Talence, France.
*vivek.kumar@lkb.ens.fr

**Abstract:** Terahertz microscopy has attracted attention owing to distinctive characteristics of the THz frequency region, particularly non-ionizing photon energy, spectral fingerprint, and transparency to most nonpolar materials. Nevertheless, the well-known Rayleigh diffraction limit imposed on THz waves commonly constrains the resultant imaging resolution to values beyond the millimeter scale, consequently limiting the applicability in numerous emerging applications for chemical sensing and complex media imaging. In this theoretical and numerical work, we address this challenge by introducing a new imaging approach, based on acquiring high-spatial frequencies by adapting the Fourier synthetic aperture approach to the terahertz spectral range, thus surpassing the diffraction-limited resolution. Our methodology combines multi-angle terahertz pulsed illumination with time-resolved field measurements, as enabled by the state-of-the-art time-domain spectroscopy technique. We demonstrate the potential of the approach for hyperspectral terahertz imaging of semi-transparent samples and show that the technique can reconstruct spatial and temporal features of complex inhomogeneous samples with subwavelength resolution.

**Keywords:** Terahertz Time-Domain Spectroscopy, Synthetic Aperture, Broadband Terahertz Imaging, Time-resolved Imaging.

**Abbreviations:** THz (terahertz), FOV (field-of-view), TDS (time-domain spectroscopy), SNR (signal-to-noise ratio), NA (numerical aperture).

1. **Introduction:**

Terahertz (THz) imaging systems are extensively used to determine the chemical and material composition of samples, marking a pivotal driving motivation in the growth of THz science as a distinct discipline [1–5]. However, the limited availability of a wide field-of-view (FOV) and high-resolution imaging systems pose a prominent technological constraint in this field of research. Specifically, at 0.3 mm wavelength (equivalent to 1 THz frequency), the far-field spatial resolution of an image is conventionally constrained to the Rayleigh criterion that corresponds to ~180 μm in vacuum, which often results in a limited space-bandwidth product [6] of the currently available THz imaging systems. To address these challenges, THz digital holography in both off-axis [7–11] and in-line [12–16] configurations have been merged with advanced computational processing algorithms such as sub-pixel sampling [14], extrapolation [17], autofocusing [15], iterative phase retrieval [14,15], and compressive sensing [16], and demonstrated improvement in the image resolution and reconstruction quality. Along the same direction, wavefront shaping techniques have recently shown promise in the THz spectral range and demonstrated single-pixel ghost imaging for complex structural objects [18–26]. However, wavefront shaping approaches involve a rather bulky and complex imaging setup due to the limited capabilities of THz wavefront modulation [27]. Similarly, ptychography imaging techniques have been introduced to accelerate high-resolution imaging without significant hardware modifications to THz imaging systems [28–30]. Quite interestingly, THz near-field imaging modalities [31–37], including tip-enhanced THz probes [35,38–40], micro-aperture THz probes [41], air-plasma dynamic aperture [42,43], have been introduced to achieve deep subwavelength resolution by capturing the object-modulated evanescent waves before the light diffraction occurs. While near-field imaging techniques have achieved deep subwavelength spatial resolution, these conventional methods are implemented through a sophisticated experimental approach and still necessitate mechanical raster scanning of an object surface with a relatively low signal-to-noise ratio (SNR), requiring a relatively long image acquisition time.

Conceptually, the resolution of an image is defined by the quantity of high-spatial frequency components present in the scattered waveform from the sample [44,45]. Moreover, during free-space propagation, high-spatial frequency components typically diffract at greater angles than low-spatial frequency components, causing the diffraction pattern to extend beyond the FOV of the detector. Consequently, a finite numerical aperture (NA) of the optical imaging system restricts the resolution by failing to capture high-spatial frequency components. In such a direction, the synthetic aperture imaging technique has been introduced to overcome the constraints imposed by the limited NA in imaging systems, which relies on capturing multiple diffraction patterns, each containing a limited spatial frequency spectrum, and coherently combining them to exploit a larger spatial frequency passband for complex samples. Considering this, synthetic aperture-based imaging was first developed at radio wavelength to improve the resolution of a single radio telescope [46,47] and subsequently adopted in optical [48–50], electron microscopy [51], sonar [52], and THz frequencies [13,53].

A significant challenge in the optical domain is the measurement of complex wavefronts, as image sensors are sensitive to the average intensity distribution of the signal. Such limited availability of field-sensitive detectors complicates the direct assessment of phase information or requires phase-retrieval algorithms [54]. In contrast, THz time-domain spectroscopy (TDS) is an established technique, able to fully resolve electric-field oscillations in very broadband pulses, thereby providing full access to the complex spectral field after the Fourier transform. A meaningful scientific question is whether the THz-TDS can be exploited to develop synthetic-aperture imaging approaches where the temporal field information can directly contribute to the high-resolution image reconstruction.

In this work, we explore this unique potential by devising a new approach for time-resolved, coherent synthetic aperture imaging. Our methodology, which outperforms the resolution limit in conventional terahertz imaging, involves synthesizing an expanded spatial frequency bandpass filter scanned across the Fourier space by illuminating the sample with a broadband THz pulse at various angles. In our imaging analysis, we then incorporate a propagation of the temporal traces obtained through TDS detection, facilitating the reconstruction of the full-field response of the sample, including phase and temporal delay contributions. This approach permits us to acquire extensive spatiotemporal information on the light-matter interaction between THz beam and the sample. After computational reconstruction, our method offers hyperspectral imaging capabilities which enable the characterization of the complex, subwavelength resolution morphology of a sample across the broadband THz spectral range.

## 2. Method
### 2.1. THz-TDS imaging system based on Fourier synthetic aperture

We present the proposed THz Fourier synthetic aperture technique implemented using a TDS imaging system. Figure 1(a) outlines our methodology, where the target object is probed by the series of THz pulses incident at several angles. The time-resolved scattered fields from the object are recorded by raster scanning a single-pixel TDS sensor on the image plane. Our imaging framework can be described as follows. A thin semi-transparent object (neglecting the absorption effect), characterized by a coherent transfer function, $\widetilde{T}(x, y; \omega)$, that we seek to recover, is illuminated by a plane wave defined as $\exp\left(i(k_{xa}x + k_{ya}y)\right)f(\omega)$, with the incident wave vectors $\vec{k}_a \equiv (k_{xa}, k_{ya})$ and $f(\omega)$ the THz broadband pulse spectrum, typically extending over a few THz. The transmitted field in its spatial frequency spectrum at a particular angular frequency $\omega$ can be expressed as,

$$\tilde{E}^+(k_x - k_{xa}, k_y - k_{ya}; \omega) = \mathcal{F}^{2D}_{x \rightarrow k}\{\tilde{T}(x, y; \omega) \cdot e^{i(k_{xa}x + k_{ya}y)}f(\omega)\} \tag{1}$$

where $\mathcal{F}^{2D}_{x \rightarrow k}\{-\}$ represents the 2D spatial Fourier transform and '·' denotes point-wise multiplication. The transmitted field spectrum, $\tilde{E}^+$, then propagates through the pupil aperture, which modulates the spatial frequency spectrum by imposing a low-pass filter. Consequently, the final spatiotemporal field at the imaging plane can be written as,

$$E_{out}(x, y; t) = \mathcal{F}^{-1D}_{\omega \rightarrow t}\left[\mathcal{F}^{-2D}_{k \rightarrow x}\{\tilde{E}^+(k_x - k_{xa}, k_y - k_{ya}; \omega) \cdot O_p^{(\omega)}(k_x, k_y)\}\right] \tag{2}$$

where $O_p^{(\omega)}$ denotes the pupil aperture due to pinhole, $\mathcal{F}_{k \to x}^{-2D}\{-\}$ represents 2D spatial inverse Fourier transform and $\mathcal{F}_{\omega \to t}^{-1D}[-]$ denotes the inverse time-Fourier transform. The overall transmitted spatiotemporal field (in Eq. (1)) at the imaging plane can be experimentally obtained using a combination of photoconductive antenna-based THz emitter and detector, readily available commercial products. Notably, the captured TDS images could immediately provide low-resolution broadband field images by performing a time-Fourier transform for each spatial point of the recorded spatiotemporal fields.

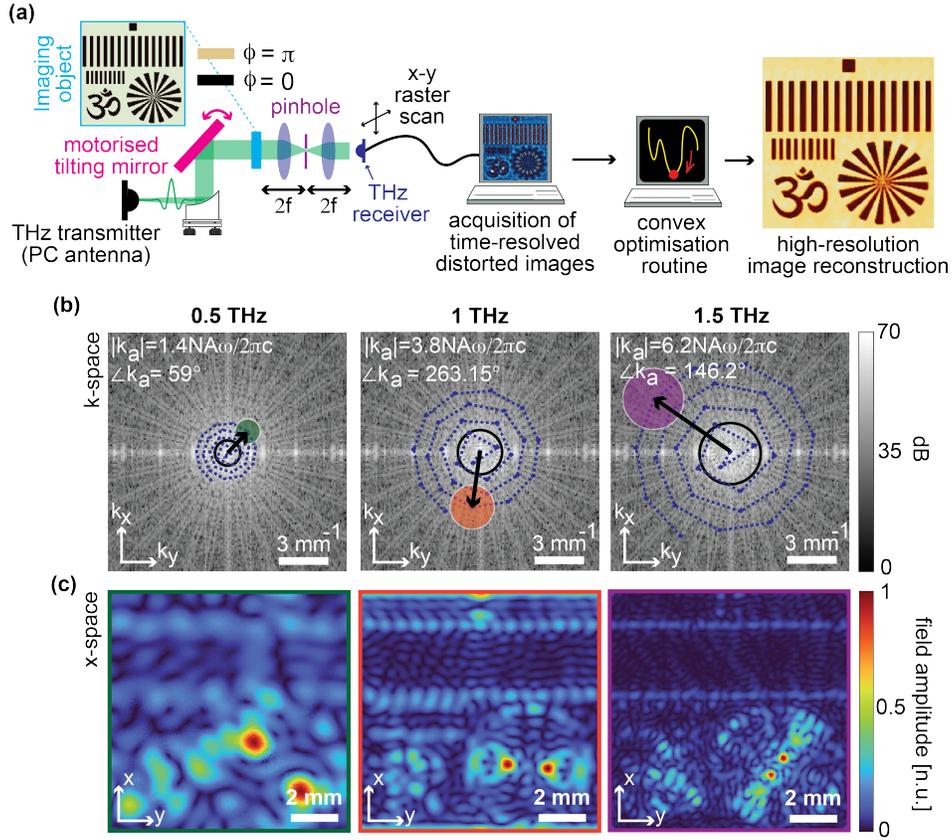

**Figure 1: Schematic of experimental-driven methodology.** (a) Conceptual overview of methodology represents illumination of the imaging object with a sequence of tilted beams of THz pulses, and corresponding time-resolved distorted images are captured using a limited aperture THz imaging system via TDS experimental approach. The spatiotemporal fields produced by the different illumination angles are formulated and processed through a convex optimisation framework that could reconstruct high-resolution images of the sample, revealing complex details previously obscured due to the finite space-bandwidth product offered by the THz imaging system. (b) Aperture synthesis in the k-space due to different tilted beam illumination defined by wavevector $\vec{k}_a \rightarrow (k_{xa}, k_{ya})$ at frequencies 0.5, 1 and 1.5 THz (Visualization 1). (c) Acquired distorted field amplitudes associated with the aperture shifts shown in (b). The $10 \times 10$ mm² object illumination area is spatially sampled at 50 μm resolution, corresponding to $200 \times 200$ pixels.

For a conventional THz imaging system operating at $\omega$ and having a finite NA, the pupil aperture is represented as a circular area formed by $k_x^2 + k_y^2 < \left(\frac{NA\omega}{2\pi c}\right)^2$ in the spatial-frequency space. This means that TDS images formed at the image plane contain spatial-frequency content limited by pupil aperture. Synthetic aperture imaging exploits the fact that the $(-k_{xa}, -k_{ya})$ wavevector offset in the transmitted field images, as shown in Eqs. (1,2) directly correlates with the illumination angle. Therefore, tweaking the illumination angle is equivalent to shifting the center of the pupil aperture, thereby enabling the capture of higher spatial frequency components that would otherwise be inaccessible within a conventional imaging framework.

For our THz-TDS proposed implementation, Fig. 1(b) illustrates how changing the illumination angles of the TDS pulses affects the shift in the spectrum of the transmitted field at 0.5, 1 and 1.5 THz, respectively. A spatial frequency low-pass filter (shaded circular region in the k-space of the object), typically controlled by a pinhole in the experimental scenarios, further constrains the measurable portion of the spectrum, resulting in a low-resolution and distorted image. A notable difference with optical implementation is that due to their very large spectral bandwidth, the effective size of the pupil aperture is impacted by the frequencies present in the THz pulse, which determines the variable aperture throughout the pulse spectrum (see Visualization 1). Figure 1(c) shows the measured field amplitude images corresponding to the spatial spectrum encompassing the pupil aperture, as shown in Fig. 1(b). It is worth noting that the hyperspectral field images, shown in Fig. 1(c), are directly acquired from the time-resolved measurements as our approach incorporates broadband THz pulse interaction with the sample.

As a next step, a set of spatiotemporal fields are recorded by varying illumination angles of the input THz pulses, and spatial frequency spectra of these transmitted fields are coherently stitched together. This implies that, by scanning the confined pupil aperture in Fourier space, following the spiral pattern shown in Fig. 1(b), we could synthesize a larger spatial frequency passband for inhomogeneous samples. To reconstruct the complex transfer function of the imaging object, we propose to implement a least-squares minimization problem that relies on reducing the error between the measured field, $\tilde{E}_{out}^a(x, y; \omega)$ and the expected field $\tilde{\xi}^a$ predicted by the forward model outlined in Eqs. (1-2). In this scenario, we define the fitness function at $\omega$ as follows:

$$\mathbb{F}(\tilde{T}_r) = \operatorname{argmin}_{\tilde{T}_r} \sum_a^{\mathcal{A}} \sum_x \sum_y \left| \tilde{E}_{out}^a(x, y; \omega) - \tilde{\xi}^a\{\tilde{T}_r(x, y; \omega)\} \right|^2 \qquad (3)$$

where 'a' corresponds to the input beam having spatial wave vector $\vec{k}_a \equiv (k_{xa}, k_{ya})$, $\mathcal{A}$ is the total number of illumination angles and $\tilde{\xi}^a\{-\}$ refers to the forward model operator that predicts the spatial field distribution on the image plane after illuminating the sample with the THz pulse of input wavevector $\vec{k}_a$. The convex optimization problem defined in Eq. (3) is further solved by employing a Nesterov accelerated gradient-descent algorithm [55], which iteratively recovers the transfer function of the imaging object. We emphasize that the fitness function described in Eq. (3) is field-sensitive, and when minimized, it can simultaneously optimize the amplitude and absolute phase of multispectral images. This gives advantages over time-averaged and intensity-based detection where access to the absolute phase with arbitrary precision is highly challenging, and waveform synthesis requires prior knowledge of the absolute phase profile of the input waveform (e.g., in the case of a transform-limited pulse).

Finally, to evaluate the reliability of our numerical approach in the presence of experimental noise, we conducted the numerical analysis under the assumption that the THz source is providing 1 nJ transform-limited pulse of duration 250 fs with 30 dB SNR measured at the detection TDS sensor. Based on this assumption, we added a white-noise component compatible with the experimental scenarios. Such relatively low noise can be readily attained in experiments with many commercial time-domain systems. Moreover, in our simulation, a 100 mm² FOV is raster scanned at 50 μm resolution by the single-pixel TDS detector, mimicking experimental implementations of the THz imaging system [56].

### 3. Results and discussion
### 3.1. Time-resolved field-sensitive imaging

To address the inherent limitations of traditional THz imaging, we implemented an image reconstruction technique that leverages the complete temporal information obtained from TDS detection. This approach yields significantly better results than conventional THz imaging techniques, which typically could operate on fixed TDS delays for image reconstruction. Furthermore, the traditional method of temporal response reconstruction, which performs reconstructions at discrete delays, needs to be improved to fully characterize the optical properties of a sample, particularly when objects are composed of complex spectral responses.

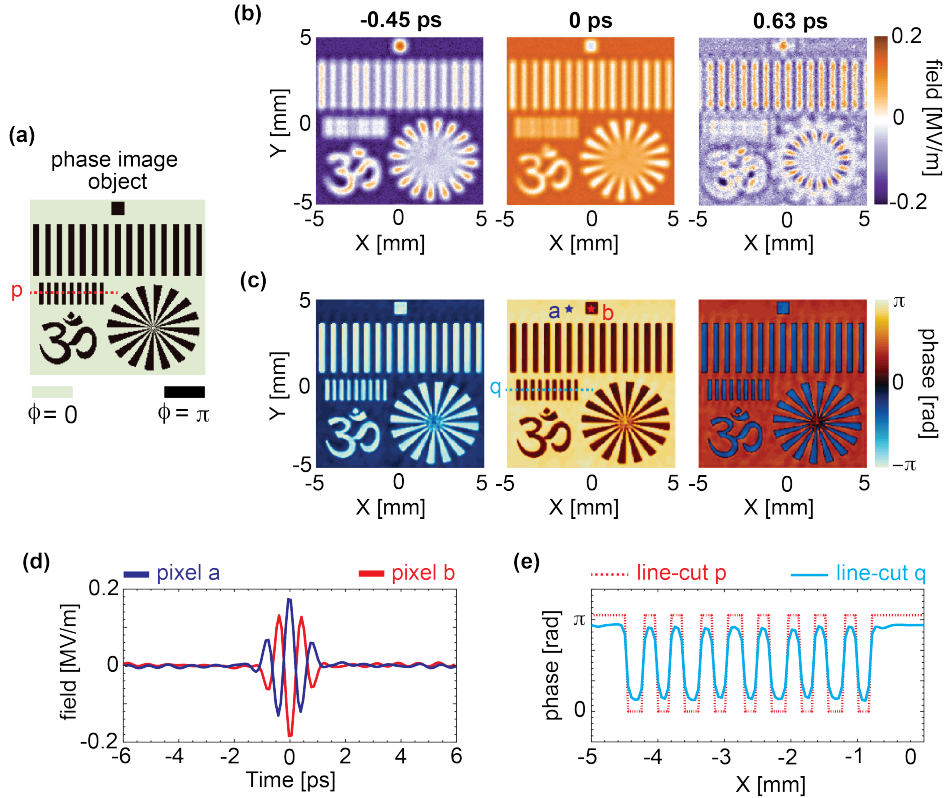

**Figure 2: Time-resolved retrieval of phase image object. (a)** Phase object under investigation. **(b)** Measured diffraction-limited, low-resolution field images at t = −0.45, 0, and 0.63 ps obtained using conventional imaging system. **(c)** Reconstructed phase images at t = −0.45, 0, and 0.63 ps. **(d)** Temporal evolution of reconstructed THz pulses for two different pixels (red and blue dots shown in panel (c)). **(e)** Line-cut plot for phase imaging object (red dashed line) and reconstructed image (cyan solid line). The 10 × 10 mm² object illumination area is spatially sampled at 50 μm resolution, corresponding to 200 × 200 pixels. We considered a 1 nJ THz pulse of duration 250 fs at the input with 30 dB SNR per pixel.

To illustrate this point, we integrated our approach in the spectral domain; in practice, the time-Fourier transform of the recorded TDS image is performed for each pixel, followed by solving the optimization problem defined in Eq. (3). After solving for each frequency $\omega$, the immediate inverse time-Fourier transform of each spatial points of hyperspectral images would result in the time-resolved reconstructed images of the sample. Figure 2 demonstrates the capabilities of our full-field methodology for achieving spatiotemporal high-resolution image retrieval, particularly for phase-sensitive reconstruction of samples with complex phase profiles. As an initial demonstration, a phase screen (Fig. 2(a)) modulates the incident field, which is further recorded using TDS detection at the image plane, resulting in diffraction-limited, low-resolution spatiotemporal images (from Eq. (2)). Figure 2(b) shows the distorted images recorded at three different times, t = -0.45, 0, and 0.63 ps, relative to our time reference of the pulse. In contrast, the image retrieval outcomes, shown in Fig. 2(c), represent the reconstructed phase images of the transmitted field at the same time values, highlighting the high fidelity of our time-resolved image reconstruction technique. After processing, the waveforms corresponding to two distinct pixels (marked by blue and red stars in panel (c)) are shown in Fig. 2(d), which are characterized by zero and $\pi$ phases of the pulses and showcase the ability to reconstruct temporal evolution of the images. Figure 2(e) illustrates the enhancement in image resolution by comparing line cuts of the phase image (dotted line in panel (a)) and the reconstructed image (dotted line in panel (c)), thereby underscoring the efficacy of our approach in improving image resolution. Interestingly, our imaging methodology can fully resolve the 150 µm slits separated by 300 µm, present in the imaging object, with marginally reduced phase contrast, which mainly depends on SNR and optical properties of the sample.

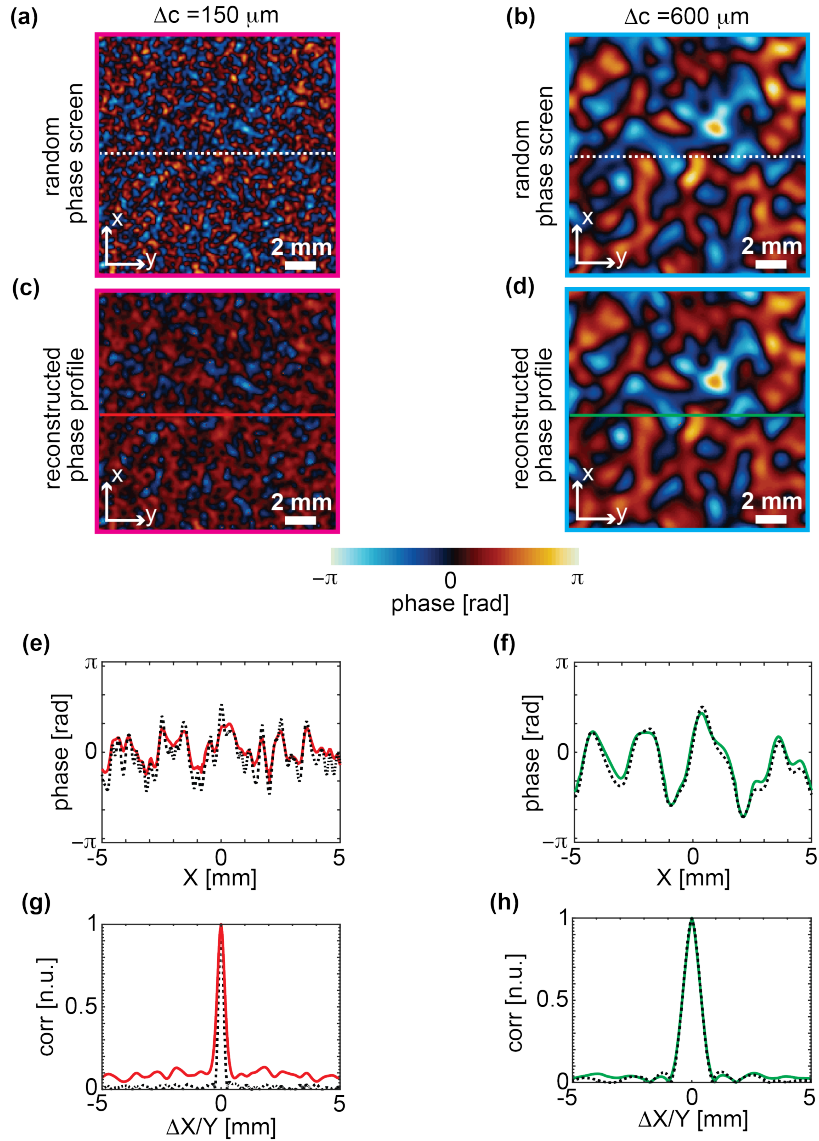

**Figure 3: Time-resolved retrieval of random media. (a, b)** Slowly varying random phase surfaces with spatial correlation-150 µm and 600 µm, respectively. **(c, d)** Retrieved spatial phase distribution of random media at t = 0 ps. **(e)** Line-cut in the x-direction for random phase screen (white dot line in panel a) and reconstructed phase image (red solid line in panel a). **(f)** line-cut in the x-direction for random phase screen (white dot line in panel b) and reconstructed phase image (green solid line in panel b). **(g,h)** Spatial autocorrelation function for the ground truth (black dashed) and retrieved (red and green) random phase screens. The $10 \times 10$ mm$^2$ object illumination area is spatially sampled at 50 µm resolution, corresponding to $200 \times 200$ pixels.

For a more challenging demonstration, our image reconstruction technique could subsequently be adapted for application to imaging in complex media. Notably, due to the larger wavelength scale of THz waves compared to optical frequencies, complex media with tunable complexity could potentially be fabricated in real experimental scenarios. Considering this, in our simulation, we decided to mimic a 3D-printed, slowly varying random surfaces that operate as passive phase-modulating elements [57], keeping the close analogy to the experiments. To model the transmission properties of such a random surface, we define a complex transfer function as follows:

$$\tilde{T}(x, y; \omega) = a_0 \exp(i\pi(n_m(\omega) - 1)A(x, y)) \tag{4}$$

where $n_m(\omega)$ represents frequency dependent refractive index of the surface material, $a_0$ is an amplitude factor, and $A(x, y)$ is topological amplitude characterized by position-dependent relative thickness distributed over the random surface. In our random media modeling, we assumed topological amplitude as a circular random Gaussian distribution with zero mean and standard deviation $\sigma = \frac{1}{2}$. To integrate controllable features and complexity into the phase element, we applied a Gaussian filter of width $\Delta c$ to the topological amplitude $A(x, y)$. Figure 3(a, b) presents imaging target random surfaces for $a_0 = 1$ and $n_m(\omega) = 1.55$ where we convolve white-noise distribution in $A(x, y)$ with a Gaussian filter with a standard deviation of $\Delta c = 150$ µm and 600 µm, respectively, to impose a desired spatial correlation in the transfer function of random media. A notable outcome is that the overall transmission characteristics are inherently associated with the degree of complexity, specifically the spatial correlation and amplitude range of the random media model. The corresponding reconstructions of the random media at $t = 0$ ps are shown in Fig. 3(c, d), where our employed technique effectively retrieved the phase topology of the random surfaces. We also plotted a line cut to show phase variation along the x-direction (shown as dashed and solid lines in Fig. 3(a-d)) in order to assess the accuracy of the reconstruction. This demonstrates the consistency in the spatial phase profile of the reconstructed image with the random phase screen. Additionally, we calculated the spatial correlation function for both the ground truth and the reconstructed random phase screens. The results are presented in Figs. 3(g, h), providing the score for the reconstruction fidelity.

### 3.2. Retrieval of THz imaging object concealed by random media

We extend our approach to obtain the transmission characteristics of a hidden object placed between two random media, as illustrated in Fig. 4. We modify the forward propagation model to evaluate the wave scattering and diffraction resulting from sequential interactions with the random phase screens and the propagation of waves across free space between them. We employ the Beam Propagation Method (BPM) [58–60] to evaluate the wave propagation through phase slices as,

$$\tilde{E}^{(q+1)}(x,y;\omega) = \tilde{T}^{(q+1)}(x,y;\omega) \cdot \wp_{\Delta z}^{(\omega)}\left[\tilde{E}^{(q)}(x,y;\omega)\right] \tag{5}$$

where $q = 1, \ldots, Q$ (in our case $Q = 3$) representing number of slices, $\tilde{E}^{(q+1)}$ and $\tilde{E}^{(q)}$ are exit field distribution in space corresponding to the $(q+1)^{th}$ and $(q)^{th}$ slices, respectively, $\tilde{T}^{q+1}$ is the transfer function of the $(q+1)^{th}$ slice, and $\Delta z$ is the separation between the slices. We represent $\wp_{\Delta z}^{(\omega)}[-]$ as a mathematical operator that propagates an electric field by the distance $\Delta z$, as per the angular spectrum propagation method. We define the free space propagator as,

$$\wp_{\Delta z}^{(\omega)}[-] = \mathcal{F}_{k\to x}^{2D}\left\{\exp\left(-i\Delta z\sqrt{\left(\frac{\omega}{c}\right)^2 - k_x^2 - k_y^2}\right) \cdot \mathcal{F}_{x\to k}^{2D}\{-\}\right\},$$ where $c$ is the speed of light in

free space, $(k_x, k_y)$ represents spatial-frequency coordinates. The BPM, functioning as a modified forward physical model for THz pulse propagation, computes the spatiotemporal wave field through sequential diffraction and multiple scattering assessments. Further additional propagation to the field exiting the final phase slice is constrained by the pupil aperture, resulting in the field on the image plane as,

$$\tilde{E}_{out}(x,y;\omega) = O_p^{(\omega)}\left[\wp_{-\Delta z}^{(\omega)}\left[\tilde{E}^{(s)}(x,y;\omega)\right]\right] \tag{6}$$

where $O_p^{(\omega)}[\![-]\!]$ denotes the pupil aperture opertor and $\wp_{-\Delta z}^{(\omega)}$ represents the back-propagation operator to refocus the exit wave to the physical focal plane, i.e., at the plane of the hidden imaging object, which is placed at the conjugate focus to the imaging plane of the system. We approach the image retrieval process as a least squares optimization problem where the disparity between the recorded and estimated field from the modified forward model can be minimized. In such case, we modify the fitness function for 3 slices as,

$$\mathbb{F}(\tilde{\mathbb{T}}_r) = \text{argmin}_{\{\tilde{T}_r^{(1)},\tilde{T}_r^{(2)},\tilde{T}_r^{(3)}\}} \sum_a^A \sum_x \sum_y \left|\tilde{E}^a_{out}(x,y;\omega) - \tilde{\xi}^a\{\tilde{T}_r^{(1)}(x,y;\omega), \tilde{T}_r^{(2)}(x,y;\omega), \tilde{T}_r^{(3)}(x,y;\omega)\}\right|^2 \tag{7}$$

where $\tilde{T}_r^{(q)}$ represents the retrieved transfer function of the $(q)^{th}$ slice and $\tilde{\xi}^a\{*\}$ refers to the modified forward model operation that predicts the spatial field distribution at frequency $\omega$ on the image plane after illuminating a sample object with an incident THz pulse defined by wavevector $\vec{k}_a \equiv (k_{xa}, k_{ya})$. Further, we use an iterative approach with the accelerated distributed Nesterov gradient descent algorithm [61] to solve the fitness function outlined in Eq. (7). As shown in Fig. 4(a), the angular projection of THz pulses interacts with an arrangement of thin slices comprising the random surfaces and hidden objects, and the complex exit field from the final slice is filtered by the pinhole, resulting in a distribution of scrambled spatiotemporal fields on the image plane. An example of a 3-slice configuration is shown in Fig. 4(b), where the first and last slices are random screens containing spatial correlation of 200 μm and 800 μm, respectively, and the middle slice represents a concealed imaging object acting as a test sample. In our simulations, each phase slice is separated by $\Delta z = 12.5$ mm, significantly larger than the THz pulse wavelength. As a relevant case example of recovering a hidden object with complex spatial structures, initially, we placed a fast-varying phase imaging object between the two random phase screens as a test sample, as shown in Fig. 4(b). As another instance, we placed a slowly varying random phase medium containing spatially correlated ($\Delta c = 300$ μm) phase elements in a circular random Gaussian distribution, as shown in Fig. 4(c). The resulting reconstruction outcomes are presented in Fig. 4(d), demonstrating the successful retrieval of each slice at $t = 0$ ps, as per the time reference of the THz pulse. The reconstruction result is also shown in Fig. 4(e) for the case of slowly varying random media. Further, we analyzed the statistical characteristics of the retrieved object, as illustrated in Fig. 4(f), by calculating the probability density function for the phase elements, showing a remarkable resemblance with the ground truth object. Moreover, we evaluated the spatial correlation embedded within the topological amplitude of the random phase screen. In Fig. 4(g), we show that the retrieved phase screen exhibits a 300 µm spatial correlation, aligning with the ground truth object. It is essential to highlight that the field-sensitive image retrieval method is effective since each slice is reconstructed simultaneously with adequate reconstruction accuracy.

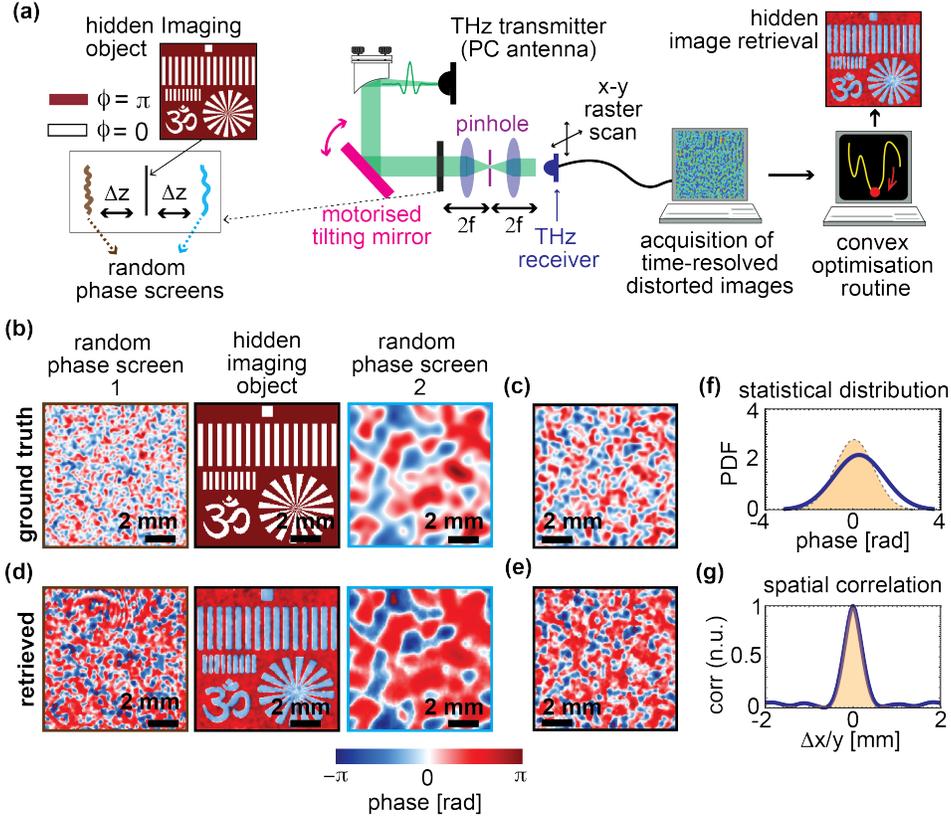

**Figure 4: Retrieval of imaging object concealed by random media. (a)** A schematic of an experimental-driven THz imaging system, where an imaging object is placed in the middle of the random phase screens separated by 25 mm. The wave interaction with random surfaces and hidden object leads to diffraction and scattering, which could be modelled as free-space wave propagation and wave scattering, respectively. The impinging wavefronts illuminate the same area of phase objects and provide redundant information in the resulting output waveform, further enabling the simultaneous reconstruction of the transmission properties of the random phase screen and the hidden objects. **(b)** Hidden object under examination with random phase surfaces containing a spatial correlation of 200 µm and 600 µm, respectively. **(c)** Hidden random phase screen with a spatial correlation of 300 µm. **(d)** Reconstructed phase profiles of the random surfaces and the hidden imaging object at t = 0 ps. **(e)** Retrieved hidden random phase screen. **(f)** Histogram (Probability density function) for the phase element of ground truth (yellow shaded area) and retrieved (blue) the random phase screen. **(g)** Spatial autocorrelation function for the ground truth (yellow shaded area) and retrieved (blue) random phase screens. The $10 \times 10$ mm$^2$ object illumination area is spatially sampled at 50 µm resolution, corresponding to $200 \times 200$ pixels.

### 3.3. Hyperspectral imaging for material characterization

As a final validation, we demonstrate the capabilities of our time-resolved imaging approach for the material characterization of complex inhomogeneous objects. We show that time-resolved image reconstruction can strongly improve hyperspectral imaging, enabling us to acquire local spectral information at each spatial coordinate of the reconstructed image. Such measurements would be crucial for spectral analysis in complex semi-transparent samples, currently presented as a significantly challenging task in conventional broadband THz imaging modalities. To explore this possibility, we simulated a semi-transparent imaging target with thickness $\Delta = 100\ \mu m$ composed of Teflon, Topas, and HDPE, as shown in Fig. 5(a). These materials exhibit markedly different spectral responses at THz frequencies, particularly in the 0.5-1.5 THz range. Therefore, selecting these materials allows direct differentiation of the spectral components and their spatial distribution within the sample. In terms of material composition and taking into account the material dispersion, the transfer function for the thin object can be expressed in terms of Fabry-Perot response as,

$$\tilde{T}(x,y;\omega) = \frac{t(x,y;\omega)e^{-i\phi(x,y;\omega)}}{1 - r(x,y;\omega)e^{-2i\phi(x,y;\omega)}} \tag{8}$$

where $\phi(x,y;\omega) = \frac{\tilde{n}(x,y;\omega)\omega\Delta}{c}$ is the function of local refractive distribution $\tilde{n}(x,y;\omega) = n(x,y;\omega) + i\rho(x,y;\omega)$, $t(x,y;\omega) = \frac{4\tilde{n}(x,y;\omega)}{[\tilde{n}(x,y;\omega)+1]^2}$ is transmission coefficient and $r(x,y;\omega) =$ $\frac{[1-\tilde{n}(x,y;\omega)]^2}{[\tilde{n}(x,y;\omega)+1]^2}$ is reflection coefficient. The results of our simulations are shown in Fig. 5. Initially, we obtained the spatial refractive index profiles under diffraction-limited imaging conditions. As shown in Fig. 5(b), the spatial refractive index distribution of the sample at 0.5, 1, 1.5 THz retrieved by conventional imaging results in distorted images, where the differentiation between materials is obscured. Conversely, as shown in the reconstructed hyperspectral image in Fig. 5(c), our method provides high-resolution spatial features in the refractive index profile with improved material distinguishability within the sample. Such enhanced material selectivity is further validated in Fig. 5(d), where we compare the reconstructed refractive indices with the material by analyzing spectral response within 0.5-1.5 THz and obtaining a satisfying agreement over a broad spectral range.

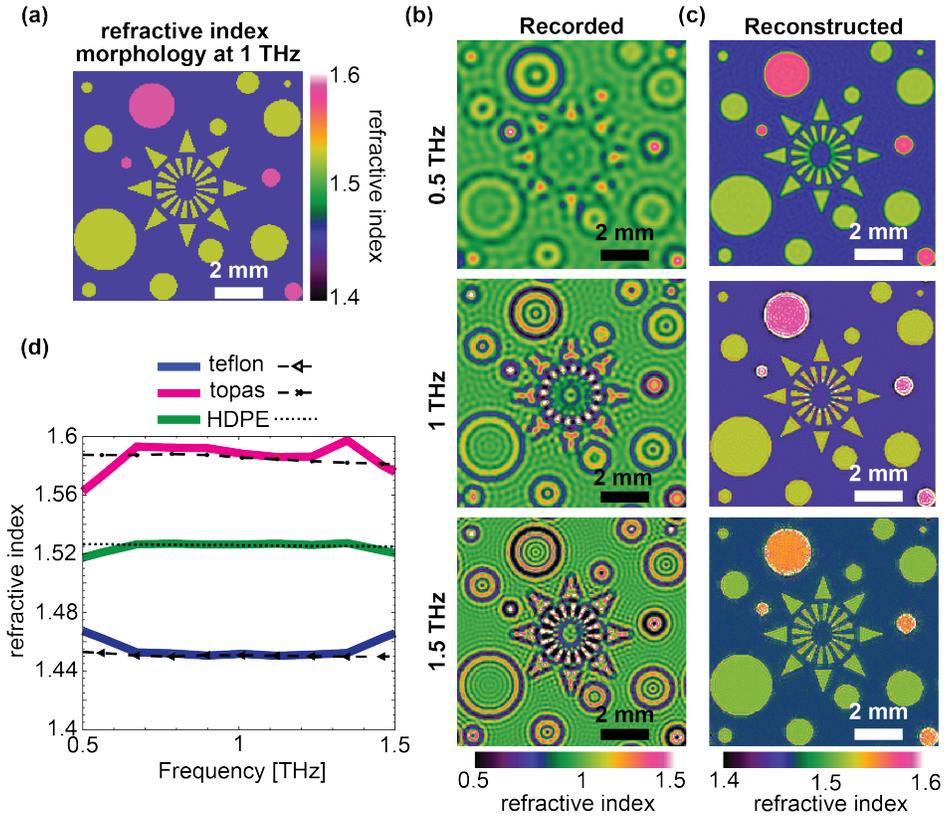

**Figure 5: Hyperspectral imaging for material characterisation. (a)** The schematic of the $100\ \mu m$ thick imaging object composed of Teflon, Topas and HDPE, showing the spatial distribution of the refractive index at 1 THz. **(b)** Distorted spatial variation in the refractive index profile at 0.5, 1 and 1.5 THz recorded using a diffraction-limited imaging system. **(c)** Reconstructed spatial refractive index profiles at 0.5, 1 and 1.5 THz illustrating the high-resolution material selectivity. **(d)** Variation in the refractive index of Teflon, Topos, and HDPE as a function of frequency representing comparison in the reconstructed refractive indices (solid lines) and material refractive indices (dashed lines). The $10 \times 10\ mm^2$ object illumination area is spatially sampled at $50\ \mu m$ resolution, corresponding to $200 \times 200$ pixles.

## 4. Conclusion

In this work, we have theoretically demonstrated a technique for time-resolved THz imaging based on Fourier synthetic aperture to overcome the limitations of traditional THz imaging methods without requiring major hardware modifications. Our methodology combines a multi-angle plane wave illumination approach and a field-sensitive THz field detection enabled by state-of-the-art TDS technology. We have demonstrated that incorporating dynamic aperture synthesis within a coherent imaging framework can achieve subwavelength resolution and enable phase-sensitive reconstruction of complex, non-dispersive samples. We also have shown image retrieval even in more complex scenarios where a phase image object is hidden between two random diffusers. Our field-sensitive optimization approach is robust, effective and compatible with experimental conditions. As a relevant case example, we demonstrated broadband imaging, successfully retrieving spatially resolved refractive index profiles for inhomogeneous and semi-transparent samples. Currently, the image reconstruction algorithm employs a convex optimization framework. While effective, it can be computationally intensive, particularly when processing large datasets or multiple angular illuminations. Algorithmic enhancements - such as more efficient solvers or parallel processing - could be explored to reduce computation time. The proposed method relies on multiple pulse illuminations from various angles to synthesize the aperture in Fourier space. Reducing the number of required illuminations without compromising resolution remains a challenge. Potential improvements could involve adaptive illumination strategies, which would select only the most informative angles, thereby reducing the overall number of illuminations needed for high-quality image reconstruction. Such an imaging approach offers novel insights into the optical properties and structural complexity of arbitrary samples, paving the way for new research and development opportunities across various fields, including chemical sensing, biomedical imaging, and material science using THz radiation.

**Funding:** This project received funding from the HYPSTER project under the ANR within the France 2030 framework: ANR-22-CE42-0005 and ANR 22-PEEL-0003 Comptera.

**Supplementary materials**
The Supporting Information is available at [LINK].
The datasets for all figures are freely accessible at: [LINK-figshare]
Additional video for aperture synthesis in the Fourier space and corresponding image formation (Visualization 1).

**Author Contributions**
V.K. conceived the idea. All authors were engaged in the discussion regarding the basic concept of the paper and its implementation for THz waves. V.K. performed the calculations and drafted the paper, with contribution from all authors on the interpretation of the results. P.Mo. and S.G. supervised the project.